\documentclass[%
reprint,
amsmath,amssymb,
]{revtex4-2}
\usepackage{graphicx}% Include figure files
\usepackage{bm}% bold math
\usepackage{url}
\newcommand{\be}{\begin{equation}}
\newcommand{\ee}{\end{equation}}
\newcommand{\bea}{\begin{eqnarray}}
\newcommand{\eea}{\end{eqnarray}}
\usepackage{bm}
\begin{document}
\title{Microwave flux-flow Hall effect in a multi-band superconductor FeSe}
\author{R. Ogawa, F. Nabeshima and A. Maeda}
\address{Department of Basic Science, The University of Tokyo, 3-8-1 Komaba, Meguro-ku, Tokyo 153-8902, Japan}

\begin{abstract}
We have measured the flux-flow Hall effect in a multi-band superconductor FeSe pure single crystal to investigate the nature of the vortex core state by means of the cross-shaped bimodal cavity technique.
We found that the flux-flow Hall angle of FeSe  at low temperatures is about 0.5, which is equal to or smaller than that evaluated by the effective viscous drag coefficient measurements.
This feature is in contrast to the cuprate superconductors.
The conductivity tensor of multi-band superconductors that are contributed from holes and electrons shows  partial cancellation of the flux-flow Hall voltage by the electrons and holes, wchich can explain the observed feature.
Therefore, our study suggests the appearance of the multi-band nature in the vortex dynamics.
\end{abstract}

\maketitle

\section{Introduction}
Quasiparticles (QPs) are confined and form quantized energy levels in a vortex core of a superconductor~\cite{Caroli1964}.
The energies of the quantized levels are expressed as $E_N=\Delta E(N+1/2)$, where $\Delta E\equiv\hbar\omega_0=\hbar \Delta_0^2/E_F$ ($\Delta_0$ and $E_F$ are the superconducting gap and Fermi energy) is the level spacing, and $N$ is an integer.
Because each level has some width $\delta E=\hbar/\tau$, the ratio of the energy spacing to its width, $r\equiv\Delta E/\delta E=\omega_0\tau$, represents the degree of quantized nature of the vortex core.
In the limit $r\gg 1$, the quantized nature inside the core is distinct, while in the opposite limit $r\ll 1$, the core can be regarded as a normal metal.
For an $r$ value between the two limits, the vortex core is called moderately clean, where the quantized nature is marginal.
As the quantized core is expected to represent many novel features, it is extremely important to know the $r$ value. 
In fact, parameter $r$ is deeply correlated with the dynamics of vortices.
Based on the conventional understanding~\cite{Blatter1994}, when an external driving current $\bm{J}=(J_0,0)$ is applied in the $x$-$y$ plane to a superconductor under a magnetic field in the $z$ direction, the driving force $\Phi_0\bm{J}\times\hat{z}$ forces a vortex, whose displacement is $\bm{u} = (x, y)$, to move in a certain direction with an angle $\phi=\tan^{-1}(\dot{y}/\dot{x}$), generating an electric field $\bm{E}$, where $\Phi_0=h/2e$ is the flux quantum, and $\hat{z}$ is a unit vector in the $z$ direction.
When we can neglect the influence of pinning, the equation of motion of a vortex is as follows:
\begin{equation}
\Phi_0\bm{J}\times\hat{z}=\eta \dot{\bm{u}}+\alpha_H\dot{\bm{u}}\times\hat{z},
\end{equation}
where $\eta=\pi\hbar nr/(1+r^2)$ is the viscous drag coefficient in the longitudinal direction, $\alpha_H=\pi\hbar nr^2/(1+r^2)$ is the viscous drag coefficient in the transvers direction, and $n$ is the density of QPs.
Solving the equation of motion yields the flux-flow Hall angle: 
\be
\tan\theta\equiv\tan(\phi-\frac{\pi}{2})=\frac{\alpha_H}{\eta}=r.
\ee
On the other hand, under a constant current, the flux-flow resistivity $\rho_{f}$ relates to the effective viscous drag coefficient as $\eta_{eff}
=\Phi_0B/ \rho_{f}~$\cite{Golosovsky1996, Tsuchiya2001}, and 
$r$ can also be represented in terms of $\eta_{eff}$ as
\be
\eta_{eff}\equiv\eta\left(1+\frac{\alpha_H^2}{\eta^2}\right)=\pi\hbar n r.
\ee

Both the measurements of the flux-flow resistivity $\rho_{f}$ and flux-flow Hall angle $\theta$ can provide information about the QPs state in the vortex core.
Thus, we can choose either of the two methods: (1) the flux-flow Hall angle measurement or (2) the effective viscous drag coefficient measurement.
So far, the latter approach has been commonly considered owing to the ease of experimentation~\cite{Klein1993,Golosovsky1994}.
To obtain information about $r$, we need the values of viscous drag coefficients at sufficiently low temperatures, well below the superconducting transition temperature $T_c$, which is  unachievable by an ordinary DC resistivity measurement because of pinning.
Note that we measured the DC Hall resistivity of FeSe$_{x}$Te$_{1-x}$ films in the mixed state near $T_c$ and found that the sign reversal of the DC Hall voltage appears often possibly due to pinning~\cite{Ogawa2018}.
Therefore, even near $T_c$, the effect of pinning cannot be neglected in DC measurements.
Thus, high frequency measurements (typically microwave)~\cite{Gittleman1968} and/or the analysis of the data by a reliable model that can obtain the value of viscous drag coefficients from the measured data~\cite{Coffey1991} are necessary.

High-$T_c$ cuprates are among the most promising candidate materials for the quantized core due to their large energy gap and small Fermi energy.  
Thus, we investigated flux flow by using  microwaves and obtained several interesting but puzzling features.
In the $\eta_{eff}$ measurement, it was found that the core in motion was moderately clean ($\omega_0\tau\sim0.1-0.3$) for a wide range of materials, independent of the cleanness of the core~\cite{Tsuchiya2001,Hanaguri1999,Maeda2007,Maeda2007a}.
However, during the direct Hall angle measurement using the microwave techniques  developed by us~\cite{Ogawa2021}, we found that $\omega_0\tau\sim1-3$ for both  Bi$_2$Sr$_2$CaCu$_2$O$_y$ and YBa$_2$Cu$_3$O$_y$, which are larger by an order of magnitude than those evaluated using the $\eta_{eff}$ measurement, and rather close to the originally expected value of $r$~\cite{Ogawa2021b}.
We explain the origin of the discrepancy in terms of the non-linearity in the viscous drag coefficient~\cite{Larkin1976a} and other possible dissipation mechanisms for the moving vortex that are not represented by the basic equation of motion of the vortex, including those recently proposed ~\cite{Hofmann1998,Hayashi1998,Smith2020a,Kogan2021a}.
However, it is not known whether such a large value and the discrepancy between the two methods will be observed in other clean superconductors.

FeSe is another promising candidate for the quantized core because of its very small Fermi energy, comparable to the superconducting gap $\Delta_0\sim E_F$~\cite{Kasahara2014}, and long QP-lifetime $\tau$ in the superconducting state~\cite{Okada2021}. %,Takahashi2011,Kurokawa2021}  
Indeed, scanning tunneling spectroscopy (STS) measurements of FeSe single crystals show Friedel-like oscillations, which shows the quantized levels in the core~\cite{Hanaguri2019}, and we expect that features specific to superclean cores will be observed.
The $\eta_{eff}$ measurement of a pure single crystal of FeSe yields   $r=1\pm0.5$~\cite{Okada2021}.
Although this is the largest value of  $r$ measured thus far among various superconductors, it is still in the moderate clean region.
We expect a much larger value in flux-flow Hall angle measurements.
Another interesting viewpoint is that FeSe is a multi-band superconductor, where we expect to observe novel effects. 
Indeed, in multi-band superconductors, the dissociation of electron and hole vortices has been  proposed theoretically~\cite{Lin2013}.
Although there have been experimental studies to investigate flux flow in multi-band superconductors~\cite{Shibata2003,Akutagawa2008,Okada2012,Takahashi2012a,Okada2013a,Okada2013b,Okada2014,Okada2015,Okada2021}, no such novel feature has been reported.

In this study, we investigated the microwave flux-flow Hall effect in FeSe pure single crystals using the cross-shaped bimodal cavity technique.
We find that the flux-flow Hall angle of FeSe at low temperatures is about 0.5, which is equal to or smaller than that evaluated by the $\eta_{eff}$ measurement.
This feature contrasts the cuprate superconductors.
We argue that it might be characteristic of multi-band superconductors  due to the cancellation of contributions from the hole and the electron.
Therefore, our study suggests the effect of the multi-band nature on the vortex dynamics.

\section{Methods}
The magnitude of the flux-flow Hall angle is given by $|\tan\theta|=|E_y/E_x|$ from Faraday's law $\bm{E}=\dot{\bm{u}}\times\bm{B}$.
Since the $z$ direction is opposite to the direction toward the interior of the sample, the definition of the surface impedance tensor  is $\tilde{Z}\equiv\bm{E}^{\parallel}(z=0)/\int^{0}_{-\infty}\bm{J}dz$, where the symbol $\parallel$ indicates that it is parallel to the surface of the sample.
This definition yields $\bm{E}\propto\tilde{Z}\bm{J}$.
Thus, the magnitude of the flux-flow Hall angle is represented as 
\begin{equation}
|\tan\theta|=\left|\frac{\sigma_{xy}}{\sigma_{xx}}\right|=\left|\frac{Z^H}{Z^L}\right|, 
\end{equation}
where $\sigma_{xx}$ and $\sigma_{xy}$ are the longitudinal and transverse conductivities, respectively, and $Z^L$ and $Z^H$ are the diagonal and off-diagonal components of the surface impedance tensor, respectively~\cite{Ogawa2021}.
Three types of measurements are necessary to obtain the magnitude of the flux-flow Hall angle:  the DC resistivity measurements, measurements of the longitudinal components of the surface impedance tensor, $Z^L$, and measurements of the transverse components of the surface impedance tensor, $Z^H$.
The DC resistivity tensor $\tilde{\rho}$ is measured using a Physical Property Measurement System (PPMS) from Quantum Design in a standard six-probe configuration. 
The longitudinal components of the surface impedances tensor $Z^L$ are measured using the conventional cavity perturbation method with a cylindrical cavity~\cite{Klein1993}, whereas
the transverse components of the surface impedance tensor $Z^H$ are measured using a cross-shaped bimodal cavity, in which the changes in the resonance characteristics and the surface impedance tensor of the sample are represented as
\begin{equation}
\Delta\left(\frac{1}{2Q}\right)=\Delta\left( G^LR^L+G^H|X^H|\right)
\end{equation}
and
\begin{equation}
\Delta\left(\frac{f}{f_0}\right)\equiv-\frac{f-f_0}{f_0}=\Delta\left( G^L X^L-G^H |R^H|\right),
\end{equation}
where $Q$ denotes the quality factor of the resonance, $f$ and $f_0$ denote the resonance frequencies with and without the sample, respectively, $R^L$ and $X^L$ are the real and imaginary parts of the longitudinal components of the surface impedance tensor ($Z^L\equiv R^L-iX^L$), respectively, $R^H$ and $X^H$ are the real and imaginary parts of the transverse components of the surface impedance tensor ($Z^H\equiv R^H-iX^H$), respectively, $G^L$ and $G^H$ represent the geometric constants in the longitudinal and transverse directions, respectively, which depend on the shape of the cavity and the sample and take the same value under the assumption that these geometries are symmetrical ($G^L=G^H$), and $\Delta$ represents the difference between the data of the same sample at different temperatures.
For explicit procedure, see ref.~\cite{Ogawa2021}.

FeSe pure single crystals were synthesized by the chemical vapor transport technique in a temperature gradient furnace using KCl/AlCl$_3$ flux~\cite{Bohmer2013}.
We performed the microwave measurements using the cross shaped bimodal cavity, operating in the two orthogonal TE$_{011}$ and TE$_{101}$ modes at 15.8 GHz, which have a quality factor $Q$ of approximately $3\times 10^3$.
The sample sizes  for the cross-shaped bimodal cavity measurements are typically $1.2\times1.2\times 0.1~ \mathrm{mm}^3$.
In all experiments, magnetic fields of up to 7 T were applied under field-cooled conditions.

\section{Results and Discussion}

Figures 1(a) and (b) show the temperature dependence of  the DC longitudinal resistivity $\rho_{xx}$ of FeSe.
$T_c^{zero}$ is about 9 K and residual resistivity ratio ($RRR=\rho_{xx}(300$~K$)/\rho_{xx}(T_{c}^{onset}=10$~K)) is about 29.
Figure 1(c) shows the magnetic field dependence of the DC Hall angle $\tan\theta_{dc}$ in the normal state. 
The DC Hall angle of FeSe single crystal above $T_c$ shows non-linearity with respect to the magnetic field due to its multi-carrier nature, as is well known~\cite{Lei2012}.
Figure 2 shows the temperature dependence of the longitudinal components of the surface impedance tensor $Z^L$ under the magnetic field.
$Z^L$ increases with an increasing magnetic field and decreases with a decreasing temperature.
This behavior is in accordance with flux-flow resistivity~\cite{Okada2021}.
Therefore, most of the observed behavior of  $Z_L$ can be attributed to flux-flow.
These results demonstrate that the sample used for experiments has the usual properties of FeSe single crystals.
Figure 3 shows the magnitude of the magnetic field dependence of the flux-flow Hall angle $|\tan\theta|$.
The uncertainty is mainly due to the small signal (change in the resonance characteristics) of the sample.
It increases with decreasing temperature due to the development of a superconducting condensate in small magnetic fields.
We observed that the magnitude of the flux-flow Hall angle at 5.3 K is typically $0.5\pm0.3$.
It is equal to or slightly smaller than that evaluated from the effective viscous drag coefficient measurement ($r= 1.0\pm 0.5$)~\cite{Okada2021}.
The flux-flow Hall angle does not depend on magnetic field strongly, even showing a very weak magnetic field dependence.
This field dependence is different from what was observed in the $\eta_{eff}$ measurement, where it increases with increasing magnetic field as $\tan\theta\sim B^{0.2}$, corresponding to the sublinear increase of the flux-flow resistivity as a function of magnetic field.
\begin{figure}[htbp]
\centering
 	\includegraphics[bb=0 0 660 420,keepaspectratio,width=86mm]{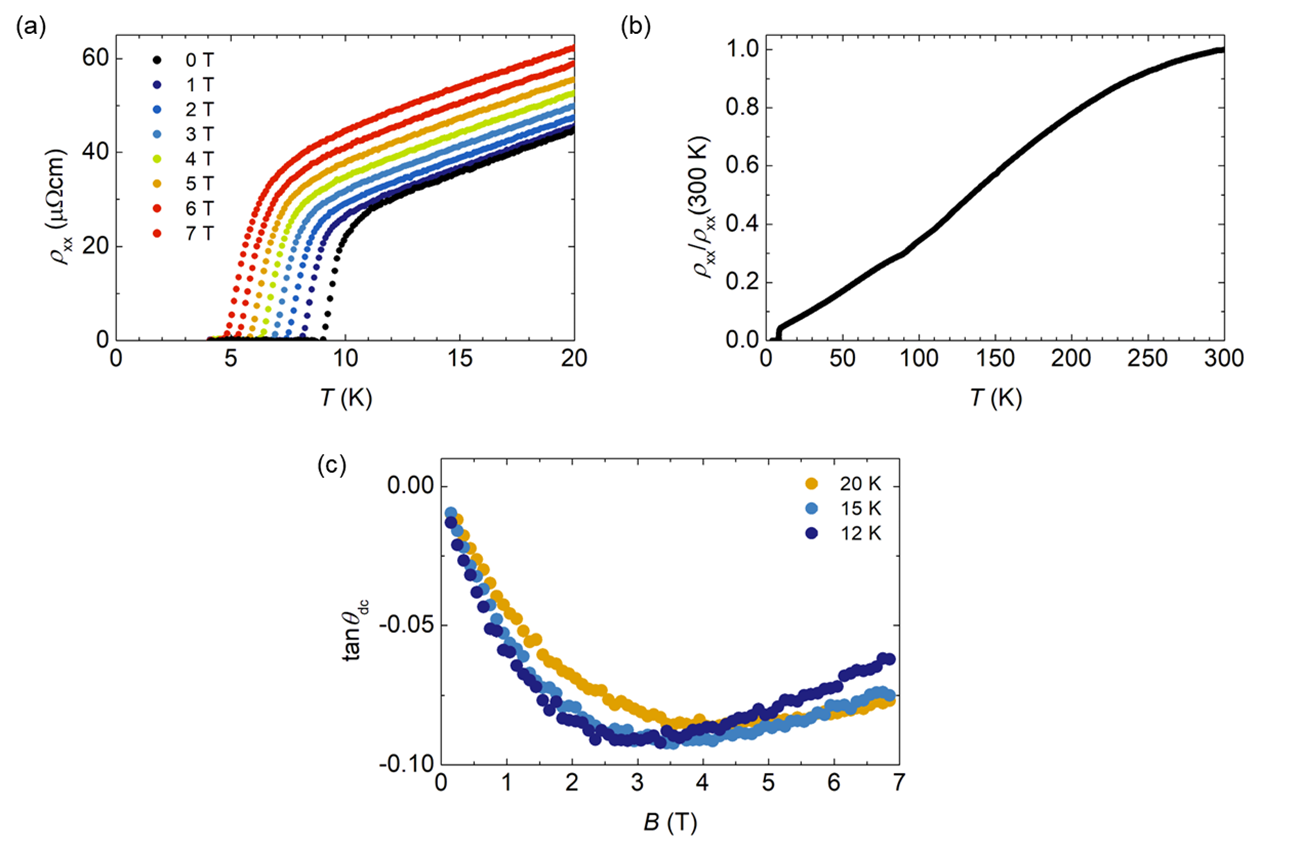}
 	\caption{ (a) The temperature dependence of the longitudinal resistivity of FeSe under a magnetic field near $T_c$ (b) The temperature dependence of the normalized longitudinal resistivity $\rho_{xx}/\rho_{xx}(300$~K$)$ at 0 T; $T_c^{zero}$ is about 9 K and $RRR=\rho _{xx}(300 $ K$)/\rho_{xx}(T_c^{onset}=10$ K$)$ is about 29 (c) The magnetic field dependence of the DC Hall angle $\tan\theta_{dc}$ in the normal state (20 K, 15 K, and 12 K, respectively). The Hall angle behaves non-monotonically with an increasing magnetic field due to its multi-carrier nature.
}
\end{figure}
\begin{figure}[htbp]
\centering
 	\includegraphics[bb=0 0 660 420, keepaspectratio,width=86mm]{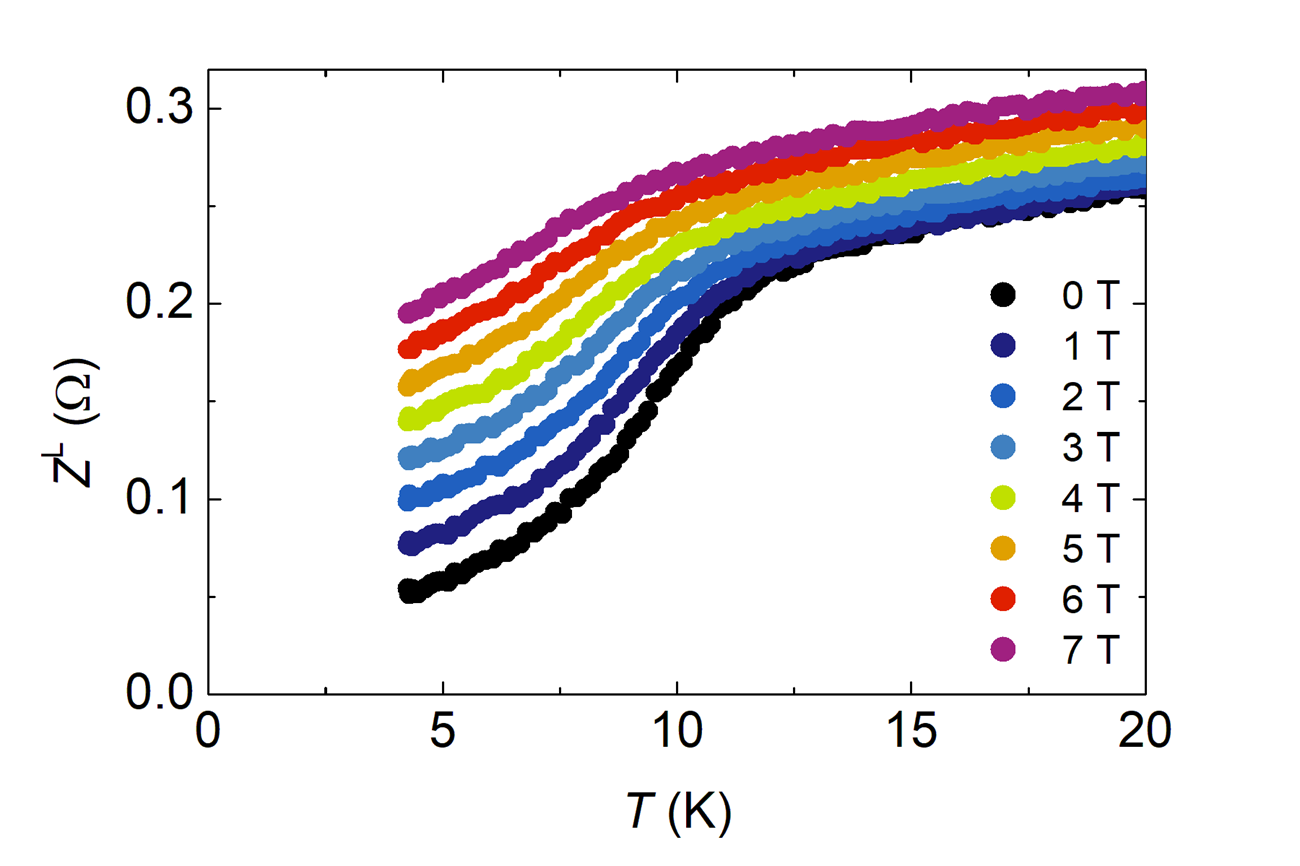}
 	\caption{Temperature dependence of the longitudinal components of the surface impedance tensor $Z^L$;  magnetic fields of up to 7 T are applied under field-cooled conditions.
}
\end{figure}
\begin{figure}[htbp]
\centering
 	\includegraphics[bb=0 0 900 600, keepaspectratio,width=86mm]{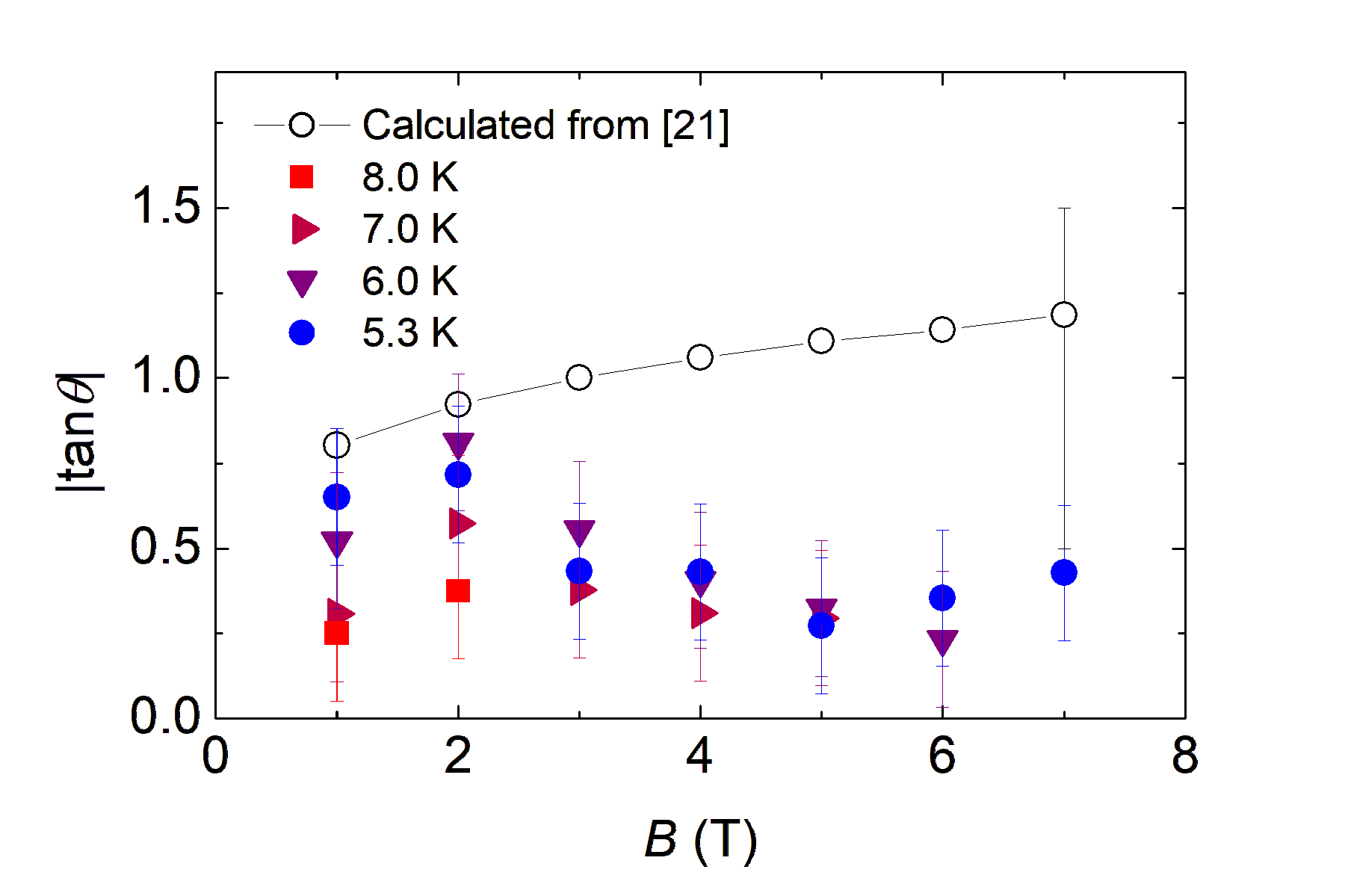}
 	\caption{Magnetic field dependence of the magnitude of the flux-flow Hall angle $|\tan\theta|$; magnetic fields of up to 7 T are applied under field-cooled conditions.
Open circles indicate the flux-flow Hall angle calculated from the data of $\eta_{eff}$ measurements~\cite{Okada2021}, whose error bar is based on $\omega_0\tau=1.0\pm0.5$ and it is written only in 7 T for ease of viewing.
The magnitude of flux-flow Hall angle  at low temperatures is equal to or smaller than that evaluated from the effective viscous drag coefficient and does not depend on magnetic-field strongly even showing a very weak field dependence.}
\end{figure}

The result of the flux-flow Hall effect in FeSe is remarkable in that the magnitude of the flux-flow Hall angle is equal to or smaller than that evaluated from $\eta_{eff}$, in contrast to cuprate superconductors.
To understand this feature, we focus on the multi-band nature of FeSe. 
FeSe has two hole bands at the $\Gamma$ point and two electron bands at the M point~\cite{Subedi2008}.
Both holes and electrons contribute to DC transport properties in the normal state, resulting in cancellation of the Hall voltage~\cite{Lei2012}.
On the other hand, no direct observation of the flux-flow Hall effect in multi-band superconductors well below $T_c$ has been reported except for DC measurements in the vicinity of $T_c$~\cite{Jin2001,Lei2010,Wang2011,Ogawa2018}.
Theoretical studies of vortex dynamics in multi-band superconductors have been carried out and the possibility of related phenomena has been pointed out~\cite{Babaev2002,Lin2013,Silaev2016}. 
In particular, a microscopic theory shows that the flux-flow conductivity tensor has contributions from both holes and electrons~\cite{Kopnin2001}.
Thus, it is possible that a similar effect as in the normal state can occur even in vortex dynamics and it explains the difference between the two methods in FeSe.
In the following we will examine the experimental results in detail from this point of view.

The microscopic theory shows the expressions for the longitudinal and transverse flux-flow conductivities at low temperatures as follows~\cite{Kopnin2001}.
\begin{eqnarray}
\sigma_{xx}=\frac{2|q|}{(2\pi)^3B}\int S(p_z)dp_z\int\frac{db}{2}\frac{\partial\epsilon_0}{\partial b}\frac{df^{(0)}(\epsilon_0)}{d\epsilon}\gamma_{xx}(\epsilon_0),
\label{sigmaxx}
\end{eqnarray}
\begin{eqnarray}
\sigma_{xy}=\frac{2q}{(2\pi)^3B}\int S(p_z)dp_z\int\frac{db}{2}\frac{\partial\epsilon_0}{\partial b}\frac{df^{(0)}(\epsilon_0)}{d\epsilon}\gamma_{xy}(\epsilon_0),
\label{sigmaxy}
\end{eqnarray}
where  $q$ denotes the charge of quasi-particles, $p_z$ denotes the momentum along the $z$ axis, $S(p_z)$ denotes the area of the cross section of the Fermi surface cut by the plane $p_z=const$, $b$ denotes the impact factor, $\epsilon_0$ denotes the energy spectrum, $f^{(0)}$ denotes the equilibrium distribution function, and $\gamma_{xx(xy)}$ are the factors defined by $\omega_0$ and $\tau$, which are an energy-interval spacing and relaxation time, as $\gamma_{xx}(\epsilon_0)=\omega_0\tau/(1+\omega_0^2\tau^2)$ and $\gamma_{xy}(\epsilon_0)=\omega_0^2\tau^2/(1+\omega_0^2\tau^2)$, respectively.

Therefore, for the single-band superconductors, the flux-flow conductivities can be obtained as a function of the quantization degrees $\omega_0\tau$.
\begin{eqnarray}
\sigma_{xx}\simeq K_{xx}\frac{\omega_0\tau}{1+(\omega_0\tau)^2},
\end{eqnarray}
\begin{eqnarray}
\sigma_{xy}\simeq K_{xy}\frac{(\omega_0\tau)^2}{1+(\omega_0\tau)^2},
\end{eqnarray}
where $K_{xx}$ and $K_{xy}$ represent the parts of conductivities other than $\omega_0\tau$.
For example, when $B\Phi_0/\pi\hbar n=K_{xx}=K_{xy}=1$, the quantization degrees experimentally evaluated by the $\tan\theta$ measurement $r_{H}$ gives the same value as that by $\eta_{eff}$ measurement $r_{\eta}$, that is, $r_H=r_{\eta}=\omega_0\tau$, since $r_{H}$ is expressed as $r_H=|\tan\theta|=|\sigma_{xy}/\sigma_{xx}|$ and $r_{\eta}$ is expressed as $r_{\eta}=\eta_{eff}/\pi\hbar n=B\Phi_{0}/\pi\hbar n\cdot(\sigma_{xx}^2+\sigma_{xy}^2)/\sigma_{xx}$.
As for the multi-band superconductors, in the flux-flow regime, a vortex is subject to the Lorentz force, which is caused by the external current, and the environmental force, which is the force from the surrounding environment (e.g., Magnus force, spectral force, {\it etc.}).
The environmental force acts on quasiparticles of the various bands $k$ in a vortex core such that $f_{env}=\sum_kf_k$.
Under the Lorentz force and the environmental force exerted on the quasi-particles in the core, the vortex core moves in a certain direction as a whole.
Consequently, the dissipation, i.e., the finite flux-flow conductivities resulting from the motion of the vortices, can be contributed by each of the different bands, which is expressed through the integral in equations (7) and (8).
Therefore,  the conductivity tensor for the multi-band superconductors can be obtained as a function of the quantization degrees of $k$th bands $(\omega_0\tau)_k$.
\begin{eqnarray}
\sigma_{xx}\simeq\sum_k K_{xx,k}\frac{(\omega_0\tau)_k}{1+(\omega_0\tau)_k^2},
\end{eqnarray}
\begin{eqnarray}
\sigma_{xy}\simeq\sum_k K_{xy,k}\frac{(\omega_0\tau)_k^2}{1+(\omega_0\tau)_k^2},
\end{eqnarray}
where $K_{xx,k}$ and $K_{xy,k}$ represent the parts of conductivities other than $(\omega_0\tau)_k$.
In the case where there are two bands, the two measurements can generally yield different quantization degrees as follows.
\begin{eqnarray}
&&r_{\eta}=\frac{B\Phi_0}{\pi\hbar n}\left(K_{xx,1}\frac{(\omega_0\tau)_1}{1+(\omega_0\tau)_1^2}+K_{xx,2}\frac{(\omega_0\tau)_2}{1+(\omega_0\tau)_2^2}\right)\nonumber \\
&&\times\left[1+
\left(\frac{K_{xy,1}\frac{(\omega_0\tau)_1^2}{1+(\omega_0\tau)_1^2}+K_{xy,2}\frac{(\omega_0\tau)_2^2}{1+(\omega_0\tau)_2^2}
}{
K_{xx,1}\frac{(\omega_0\tau)_1}{1+(\omega_0\tau)_1^2}+K_{xx,2}\frac{(\omega_0\tau)_2}{1+(\omega_0\tau)_2^2}
}\right)^2
\right],
\end{eqnarray}
\begin{eqnarray}
r_{H}=\left|\frac{K_{xy,1}\frac{(\omega_0\tau)_1^2}{1+(\omega_0\tau)_1^2}+K_{xy,2}\frac{(\omega_0\tau)_2^2}{1+(\omega_0\tau)_2^2}
}{
K_{xx,1}\frac{(\omega_0\tau)_1}{1+(\omega_0\tau)_1^2}+K_{xx,2}\frac{(\omega_0\tau)_2}{1+(\omega_0\tau)_2^2}
}\right|.
\end{eqnarray}
In particular, when there are hole bands $(\omega_0\tau)_1$ and electron bands $(\omega_0\tau)_2$, they cancel out each other in the transverse conductivity because the sign of $K_{xy,1}$ is opposite to that of  $K_{xy,2}$.
In Figure 4, we plot $r_H$ and $r_\eta$ as a function of $(\omega_0\tau)_2$ for some selected values of $(\omega_0\tau)_1$, where we assume that $B\Phi_0/\pi\hbar n=K_{xx,1}=K_{xx,2}=K_{xy,1}=1$ and $K_{xy,2}=-1$ for simplicity, which means that $r_H$ and $r_{\eta}$ give the same value in the single band case.
In all figures, $r_H$ is always smaller than $r_{\eta}$ except for $(\omega_0\tau)_2=0$.
Therefore, the smallness of $r_H$ in FeSe, which is different from the case of cuprate superconductors, can be interpreted by the effect of the multi-band nature.
\begin{figure}[htbp]
\centering
 	\includegraphics[bb=0 0 1150 750, keepaspectratio,width=86mm]{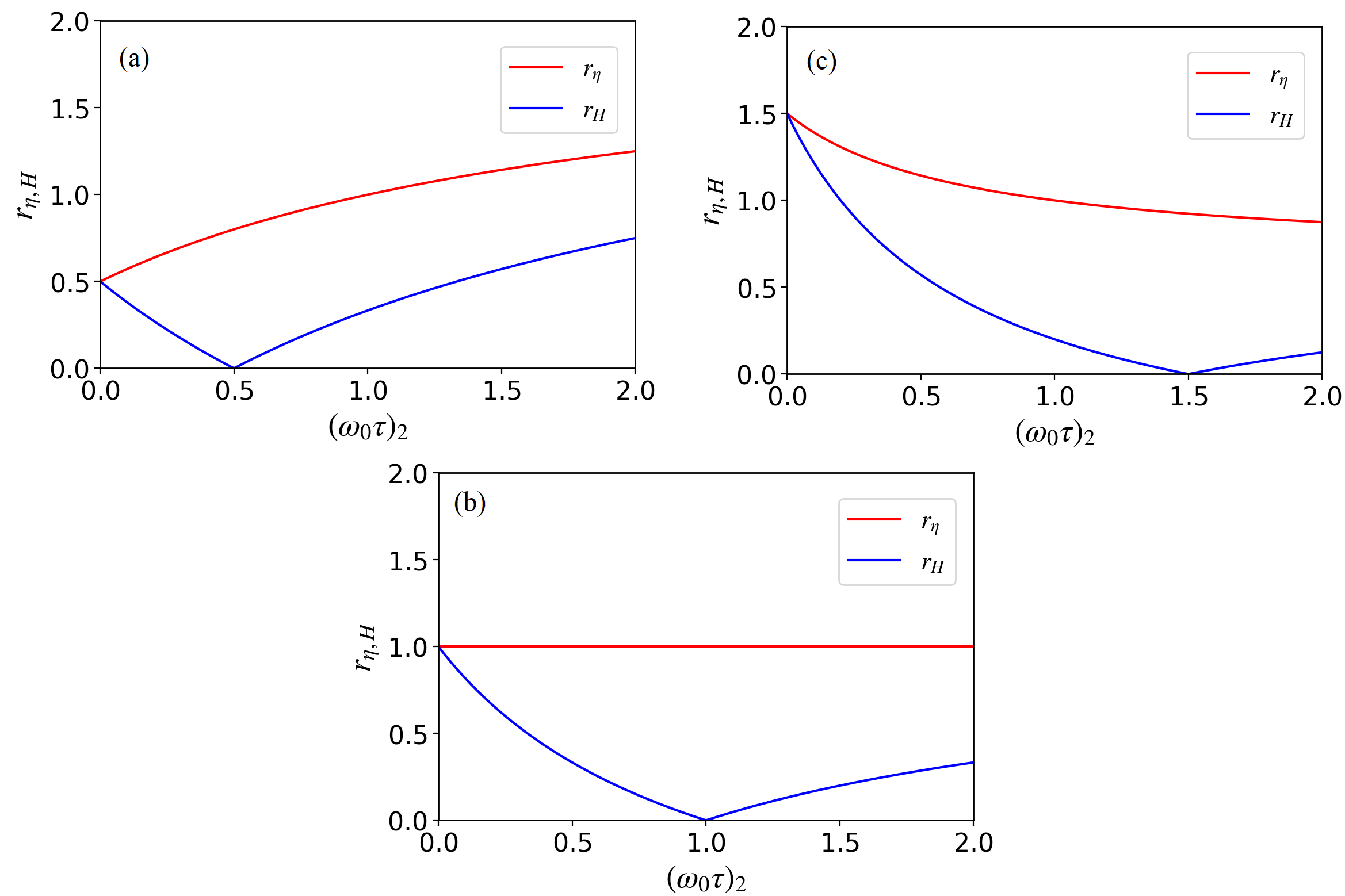}
 	\caption{The quantization degree $(\omega_0\tau)_2$ dependence of $r_H$ and $r_{\eta}$, when there are hole and electron bands for (a) $(\omega_0\tau)_1=0.5$, (b) $(\omega_0\tau)_1=1.0$ and (c) $(\omega_0\tau)_1=1.5$.
For simplicity, we assume that $B\Phi_0/n\pi\hbar=K_{xx,1}=K_{xx,2}=K_{xy,1}=1$ and $K_{xy,2}=-1$.
The blue lines represent values obtained from $\tan\theta$ measurements $\tilde{r}_H$, and the red lines represent those obtained from $\eta_{eff}$ measurements $\tilde{r}_{\eta}$. 
It shows that $r_H$ is always smaller than $r_{\eta}$ except for $(\omega_0\tau)_2=0$ (the single band case). }
\end{figure}

The difference between $r_H$ and $r_{\eta}$ is also seen in the difference in the magnetic field dependence, as shown in Figure 3.
Although the error bars in the $\eta_{eff}$ measurement are large due to the large magnetoresistance in the normal state, we can safely mention that $r_H$ and $r_{\eta}$ behave differently as a function of the magnetic field.
The Hall angle measurements show the weak magnetic field dependence as mentioned, while equation (14)  suggests that $\tan\theta=\sigma_{xy}/\sigma_{xx}$ does not show a strong magnetic field dependence.
Here, it should be noted that these equations were derived under the assumption that the order parameter is uniform (isotropic). 
Experiments have shown that the superconducting gap of FeSe is highly anisotropic ~\cite{Kasahara2014,Liu2018}. 
In Fe-based superconductors including FeSe, although equation (13) also suggests that $\eta_{eff}$ does not show a strong magnetic field dependence, the sublinear magnetic field dependence of the flux-flow resistivity is obtained by the $\eta_{eff}$ study~\cite{Okada2014,Okada2020}.
Therefore, the anisotropy of the superconducting gap may explain the behavior of the magnetic field dependence in the flux-flow Hall angle.

\section{Conclusion}
In summary, we have measured the flux-flow Hall effect in a multi-band superconductor FeSe pure single crystal to investigate the nature of the vortex core state by means of the cross-shaped bimodal cavity technique.
We found that the flux-flow Hall angle of FeSe  at low temperatures is about 0.5, which is equal to or smaller than that evaluated by the $\eta_{eff}$ measurements.
This feature is in contrast to the cuprate superconductors.
The conductivity tensor of multi-band superconductors that are contributed from holes and electrons shows  partial cancellation of the flux-flow Hall voltage by the electrons and holes, which can explain the observed feature.
Therefore, our study suggests the appearance of the multi-band nature in the vortex dynamics.

%\section*{References}


\begin{thebibliography}{40}%
\makeatletter
\providecommand \@ifxundefined [1]{%
 \@ifx{#1\undefined}
}%
\providecommand \@ifnum [1]{%
 \ifnum #1\expandafter \@firstoftwo
 \else \expandafter \@secondoftwo
 \fi
}%
\providecommand \@ifx [1]{%
 \ifx #1\expandafter \@firstoftwo
 \else \expandafter \@secondoftwo
 \fi
}%
\providecommand \natexlab [1]{#1}%
\providecommand \enquote  [1]{``#1''}%
\providecommand \bibnamefont  [1]{#1}%
\providecommand \bibfnamefont [1]{#1}%
\providecommand \citenamefont [1]{#1}%
\providecommand \href@noop [0]{\@secondoftwo}%
\providecommand \href [0]{\begingroup \@sanitize@url \@href}%
\providecommand \@href[1]{\@@startlink{#1}\@@href}%
\providecommand \@@href[1]{\endgroup#1\@@endlink}%
\providecommand \@sanitize@url [0]{\catcode `\\12\catcode `\$12\catcode
  `\&12\catcode `\#12\catcode `\^12\catcode `\_12\catcode `\%12\relax}%
\providecommand \@@startlink[1]{}%
\providecommand \@@endlink[0]{}%
\providecommand \url  [0]{\begingroup\@sanitize@url \@url }%
\providecommand \@url [1]{\endgroup\@href {#1}{\urlprefix }}%
\providecommand \urlprefix  [0]{URL }%
\providecommand \Eprint [0]{\href }%
\providecommand \doibase [0]{http://dx.doi.org/}%
\providecommand \selectlanguage [0]{\@gobble}%
\providecommand \bibinfo  [0]{\@secondoftwo}%
\providecommand \bibfield  [0]{\@secondoftwo}%
\providecommand \translation [1]{[#1]}%
\providecommand \BibitemOpen [0]{}%
\providecommand \bibitemStop [0]{}%
\providecommand \bibitemNoStop [0]{.\EOS\space}%
\providecommand \EOS [0]{\spacefactor3000\relax}%
\providecommand \BibitemShut  [1]{\csname bibitem#1\endcsname}%
\let\auto@bib@innerbib\@empty
%</preamble>
\bibitem [{\citenamefont {Caroli}\ \emph {et~al.}(1964)\citenamefont {Caroli},
  \citenamefont {{De Gennes}},\ and\ \citenamefont {Matricon}}]{Caroli1964}%
  \BibitemOpen
  \bibfield  {author} {\bibinfo {author} {\bibfnamefont {C.}~\bibnamefont
  {Caroli}}, \bibinfo {author} {\bibfnamefont {P.}~\bibnamefont {{De Gennes}}},
  \ and\ \bibinfo {author} {\bibfnamefont {J.}~\bibnamefont {Matricon}},\
  } {\bibfield  {journal} {\bibinfo  {journal}
  {Physics Letters}\ }\textbf {\bibinfo {volume} {9}},\ \bibinfo {pages} {307}
  (\bibinfo {year} {1964})}\BibitemShut {NoStop}% 
\bibitem [{\citenamefont {Blatter}\ \emph {et~al.}(1994)\citenamefont
  {Blatter}, \citenamefont {Feigel'man}, \citenamefont {Geshkenbein},
  \citenamefont {Larkin},\ and\ \citenamefont {Vinokur}}]{Blatter1994}%
  \BibitemOpen
  \bibfield  {author} {\bibinfo {author} {\bibfnamefont {G.}~\bibnamefont
  {Blatter}}, \bibinfo {author} {\bibfnamefont {M.~V.}\ \bibnamefont
  {Feigel'man}}, \bibinfo {author} {\bibfnamefont {V.~B.}\ \bibnamefont
  {Geshkenbein}}, \bibinfo {author} {\bibfnamefont {A.~I.}\ \bibnamefont
  {Larkin}}, \ and\ \bibinfo {author} {\bibfnamefont {V.~M.}\ \bibnamefont
  {Vinokur}},\ } {\bibfield  {journal} {\bibinfo  {journal}
  {Reviews of Modern Physics}\ }\textbf {\bibinfo {volume} {66}},\ \bibinfo
  {pages} {1125} (\bibinfo {year} {1994})}\BibitemShut {NoStop}%
\bibitem [{\citenamefont {Golosovsky}\ \emph {et~al.}(1996)\citenamefont
  {Golosovsky}, \citenamefont {Tsindlekht},\ and\ \citenamefont
  {Davidov}}]{Golosovsky1996}%
  \BibitemOpen
  \bibfield  {author} {\bibinfo {author} {\bibfnamefont {M.}~\bibnamefont
  {Golosovsky}}, \bibinfo {author} {\bibfnamefont {M.}~\bibnamefont
  {Tsindlekht}}, \ and\ \bibinfo {author} {\bibfnamefont {D.}~\bibnamefont
  {Davidov}},\ }{\bibfield  {journal} {\bibinfo  {journal}
  {Superconductor Science and Technology}\ }\textbf {\bibinfo {volume} {9}},\
  \bibinfo {pages} {1} (\bibinfo {year} {1996})}\BibitemShut {NoStop}%
\bibitem [{\citenamefont {Tsuchiya}\ \emph {et~al.}(2001)\citenamefont
  {Tsuchiya}, \citenamefont {Iwaya}, \citenamefont {Kinoshita}, \citenamefont
  {Hanaguri}, \citenamefont {Kitano}, \citenamefont {Maeda}, \citenamefont
  {Shibata}, \citenamefont {Nishizaki},\ and\ \citenamefont
  {Kobayashi}}]{Tsuchiya2001}%
  \BibitemOpen
  \bibfield  {author} {\bibinfo {author} {\bibfnamefont {Y.}~\bibnamefont
  {Tsuchiya}}, \bibinfo {author} {\bibfnamefont {K.}~\bibnamefont {Iwaya}},
  \bibinfo {author} {\bibfnamefont {K.}~\bibnamefont {Kinoshita}}, \bibinfo
  {author} {\bibfnamefont {T.}~\bibnamefont {Hanaguri}}, \bibinfo {author}
  {\bibfnamefont {H.}~\bibnamefont {Kitano}}, \bibinfo {author} {\bibfnamefont
  {A.}~\bibnamefont {Maeda}}, \bibinfo {author} {\bibfnamefont
  {K.}~\bibnamefont {Shibata}}, \bibinfo {author} {\bibfnamefont
  {T.}~\bibnamefont {Nishizaki}}, \ and\ \bibinfo {author} {\bibfnamefont
  {N.}~\bibnamefont {Kobayashi}},\ }{\bibfield  {journal} {\bibinfo  {journal}
  {Physical Review B}\ }\textbf {\bibinfo {volume} {63}},\ \bibinfo {pages}
  {184517} (\bibinfo {year} {2001})}\BibitemShut {NoStop}%
\bibitem [{\citenamefont {Klein}\ \emph {et~al.}(1993)\citenamefont {Klein},
  \citenamefont {Donovan}, \citenamefont {Dressel},\ and\ \citenamefont
  {Gr{\"{u}}ner}}]{Klein1993}%
  \BibitemOpen
  \bibfield  {author} {\bibinfo {author} {\bibfnamefont {O.}~\bibnamefont
  {Klein}}, \bibinfo {author} {\bibfnamefont {S.}~\bibnamefont {Donovan}},
  \bibinfo {author} {\bibfnamefont {M.}~\bibnamefont {Dressel}}, \ and\
  \bibinfo {author} {\bibfnamefont {G.}~\bibnamefont {Gr{\"{u}}ner}},\
  }
  {\bibfield  {journal} {\bibinfo  {journal} {International Journal of Infrared
  and Millimeter Waves}\ }\textbf {\bibinfo {volume} {14}},\ \bibinfo {pages}
  {2423} (\bibinfo {year} {1993})}\BibitemShut {NoStop}%
\bibitem [{\citenamefont {Golosovsky}\ \emph {et~al.}(1994)\citenamefont
  {Golosovsky}, \citenamefont {Tsindlekht}, \citenamefont {Chayet},\ and\
  \citenamefont {Davidov}}]{Golosovsky1994}%
  \BibitemOpen
  \bibfield  {author} {\bibinfo {author} {\bibfnamefont {M.}~\bibnamefont
  {Golosovsky}}, \bibinfo {author} {\bibfnamefont {M.}~\bibnamefont
  {Tsindlekht}}, \bibinfo {author} {\bibfnamefont {H.}~\bibnamefont {Chayet}},
  \ and\ \bibinfo {author} {\bibfnamefont {D.}~\bibnamefont {Davidov}},\
  }{\bibfield
  {journal} {\bibinfo  {journal} {Physical Review B}\ }\textbf {\bibinfo
  {volume} {50}},\ \bibinfo {pages} {470} (\bibinfo {year} {1994})}\BibitemShut
  {NoStop}%
\bibitem [{\citenamefont {Ogawa}\ \emph {et~al.}(2018)\citenamefont {Ogawa},
  \citenamefont {Ishikawa}, \citenamefont {Kawai}, \citenamefont {Nabeshima},\
  and\ \citenamefont {Maeda}}]{Ogawa2018}%
  \BibitemOpen
  \bibfield  {author} {\bibinfo {author} {\bibfnamefont {R.}~\bibnamefont
  {Ogawa}}, \bibinfo {author} {\bibfnamefont {T.}~\bibnamefont {Ishikawa}},
  \bibinfo {author} {\bibfnamefont {M.}~\bibnamefont {Kawai}}, \bibinfo
  {author} {\bibfnamefont {F.}~\bibnamefont {Nabeshima}}, \ and\ \bibinfo
  {author} {\bibfnamefont {A.}~\bibnamefont {Maeda}},\ } {\bibfield  {journal} {\bibinfo  {journal}
  {Journal of Physics: Conference Series}\ }\textbf {\bibinfo {volume}
  {1054}},\ \bibinfo {pages} {012021} (\bibinfo {year} {2018})}\BibitemShut
  {NoStop}%
\bibitem [{\citenamefont {Gittleman}\ and\ \citenamefont
  {Rosenblum}(1968)}]{Gittleman1968}%
  \BibitemOpen
  \bibfield  {author} {\bibinfo {author} {\bibfnamefont {J.~I.}\ \bibnamefont
  {Gittleman}}\ and\ \bibinfo {author} {\bibfnamefont {B.}~\bibnamefont
  {Rosenblum}},\ } {\bibfield  {journal} {\bibinfo  {journal}
  {Journal of Applied Physics}\ }\textbf {\bibinfo {volume} {39}},\ \bibinfo
  {pages} {2617} (\bibinfo {year} {1968})}\BibitemShut {NoStop}%
\bibitem [{\citenamefont {Coffey}\ and\ \citenamefont
  {Clem}(1991)}]{Coffey1991}%
  \BibitemOpen
  \bibfield  {author} {\bibinfo {author} {\bibfnamefont {M.~W.}\ \bibnamefont
  {Coffey}}\ and\ \bibinfo {author} {\bibfnamefont {J.~R.}\ \bibnamefont
  {Clem}},\ }
  {\bibfield  {journal} {\bibinfo  {journal} {Physical Review Letters}\
  }\textbf {\bibinfo {volume} {67}},\ \bibinfo {pages} {386} (\bibinfo {year}
  {1991})}\BibitemShut {NoStop}%
\bibitem [{\citenamefont {Hanaguri}\ \emph {et~al.}(1999)\citenamefont
  {Hanaguri}, \citenamefont {Tsuboi}, \citenamefont {Tsuchiya}, \citenamefont
  {Sasaki},\ and\ \citenamefont {Maeda}}]{Hanaguri1999}%
  \BibitemOpen
  \bibfield  {author} {\bibinfo {author} {\bibfnamefont {T.}~\bibnamefont
  {Hanaguri}}, \bibinfo {author} {\bibfnamefont {T.}~\bibnamefont {Tsuboi}},
  \bibinfo {author} {\bibfnamefont {Y.}~\bibnamefont {Tsuchiya}}, \bibinfo
  {author} {\bibfnamefont {K.}~\bibnamefont {Sasaki}}, \ and\ \bibinfo {author}
  {\bibfnamefont {A.}~\bibnamefont {Maeda}},\ }{\bibfield  {journal} {\bibinfo  {journal}
  {Physical Review Letters}\ }\textbf {\bibinfo {volume} {82}},\ \bibinfo
  {pages} {1273} (\bibinfo {year} {1999})}\BibitemShut {NoStop}%
\bibitem [{\citenamefont {Maeda}\ \emph
  {et~al.}(2007{\natexlab{a}})\citenamefont {Maeda}, \citenamefont {Kitano},
  \citenamefont {Kinoshita}, \citenamefont {Nishizaki}, \citenamefont
  {Shibata},\ and\ \citenamefont {Kobayashi}}]{Maeda2007}%
  \BibitemOpen
  \bibfield  {author} {\bibinfo {author} {\bibfnamefont {A.}~\bibnamefont
  {Maeda}}, \bibinfo {author} {\bibfnamefont {H.}~\bibnamefont {Kitano}},
  \bibinfo {author} {\bibfnamefont {K.}~\bibnamefont {Kinoshita}}, \bibinfo
  {author} {\bibfnamefont {T.}~\bibnamefont {Nishizaki}}, \bibinfo {author}
  {\bibfnamefont {K.}~\bibnamefont {Shibata}}, \ and\ \bibinfo {author}
  {\bibfnamefont {N.}~\bibnamefont {Kobayashi}},\ }{\bibfield  {journal} {\bibinfo  {journal} {Journal
  of the Physical Society of Japan}\ }\textbf {\bibinfo {volume} {76}},\
  \bibinfo {pages} {094708} (\bibinfo {year} {2007}{\natexlab{a}})}\BibitemShut
  {NoStop}%
\bibitem [{\citenamefont {Maeda}\ \emph
  {et~al.}(2007{\natexlab{b}})\citenamefont {Maeda}, \citenamefont {Umetsu},\
  and\ \citenamefont {Kitano}}]{Maeda2007a}%
  \BibitemOpen
  \bibfield  {author} {\bibinfo {author} {\bibfnamefont {A.}~\bibnamefont
  {Maeda}}, \bibinfo {author} {\bibfnamefont {T.}~\bibnamefont {Umetsu}}, \
  and\ \bibinfo {author} {\bibfnamefont {H.}~\bibnamefont {Kitano}},\
  }
  {\bibfield  {journal} {\bibinfo  {journal} {Physica C: Superconductivity}\
  }\textbf {\bibinfo {volume} {460-462}},\ \bibinfo {pages} {1202} (\bibinfo
  {year} {2007}{\natexlab{b}})}\BibitemShut {NoStop}%
\bibitem [{\citenamefont {Ogawa}\ \emph
  {et~al.}(2021{\natexlab{a}})\citenamefont {Ogawa}, \citenamefont {Okada},
  \citenamefont {Takahashi}, \citenamefont {Nabeshima},\ and\ \citenamefont
  {Maeda}}]{Ogawa2021}%
  \BibitemOpen
  \bibfield  {author} {\bibinfo {author} {\bibfnamefont {R.}~\bibnamefont
  {Ogawa}}, \bibinfo {author} {\bibfnamefont {T.}~\bibnamefont {Okada}},
  \bibinfo {author} {\bibfnamefont {H.}~\bibnamefont {Takahashi}}, \bibinfo
  {author} {\bibfnamefont {F.}~\bibnamefont {Nabeshima}}, \ and\ \bibinfo
  {author} {\bibfnamefont {A.}~\bibnamefont {Maeda}},\ }{\bibfield
  {journal} {\bibinfo  {journal} {Journal of Applied Physics}\ }\textbf
  {\bibinfo {volume} {129}},\ \bibinfo {pages} {015102} (\bibinfo {year}
  {2021}{\natexlab{a}})} \BibitemShut {NoStop}%
\bibitem [{\citenamefont {Ogawa}\ \emph
  {et~al.}(2021{\natexlab{b}})\citenamefont {Ogawa}, \citenamefont {Nabeshima},
  \citenamefont {Nishizaki},\ and\ \citenamefont {Maeda}}]{Ogawa2021b}%
  \BibitemOpen
  \bibfield  {author} {\bibinfo {author} {\bibfnamefont {R.}~\bibnamefont
  {Ogawa}}, \bibinfo {author} {\bibfnamefont {F.}~\bibnamefont {Nabeshima}},
  \bibinfo {author} {\bibfnamefont {T.}~\bibnamefont {Nishizaki}}, \ and\
  \bibinfo {author} {\bibfnamefont {A.}~\bibnamefont {Maeda}},\ }{\bibfield
  {journal} {\bibinfo  {journal} {Physical Review B}\ }\textbf {\bibinfo
  {volume} {104}},\ \bibinfo {pages} {L020503} (\bibinfo {year}
  {2021}{\natexlab{b}})}\BibitemShut {NoStop}%
\bibitem [{\citenamefont {Larkin}\ and\ \citenamefont
  {Ovchinnikov}(1976)}]{Larkin1976a}%
  \BibitemOpen
  \bibfield  {author} {\bibinfo {author} {\bibfnamefont {A.}~\bibnamefont
  {Larkin}}\ and\ \bibinfo {author} {\bibfnamefont {Y.}~\bibnamefont
  {Ovchinnikov}},\ }
  {\bibfield  {journal} {\bibinfo  {journal} {JETP}\ }\textbf {\bibinfo
  {volume} {41}},\ \bibinfo {pages} {960} (\bibinfo {year} {1976})}\BibitemShut
  {NoStop}%
\bibitem [{\citenamefont {Hofmann}\ and\ \citenamefont
  {K{\"{u}}mmel}(1998)}]{Hofmann1998}%
  \BibitemOpen
  \bibfield  {author} {\bibinfo {author} {\bibfnamefont {S.}~\bibnamefont
  {Hofmann}}\ and\ \bibinfo {author} {\bibfnamefont {R.}~\bibnamefont
  {K{\"{u}}mmel}},\ } {\bibfield  {journal} {\bibinfo
  {journal} {Physical Review B}\ }\textbf {\bibinfo {volume} {57}},\ \bibinfo
  {pages} {7904} (\bibinfo {year} {1998})}\BibitemShut {NoStop}%
\bibitem [{\citenamefont {Hayashi}(1998)}]{Hayashi1998}%
  \BibitemOpen
  \bibfield  {author} {\bibinfo {author} {\bibfnamefont {M.}~\bibnamefont
  {Hayashi}},\ } {\bibfield  {journal} {\bibinfo
  {journal} {Journal of the Physical Society of Japan}\ }\textbf {\bibinfo
  {volume} {67}},\ \bibinfo {pages} {3372} (\bibinfo {year}
  {1998})}\BibitemShut {NoStop}%
\bibitem [{\citenamefont {Smith}\ \emph {et~al.}(2020)\citenamefont {Smith},
  \citenamefont {Andreev}, \citenamefont {Feigel'man},\ and\ \citenamefont
  {Spivak}}]{Smith2020a}%
  \BibitemOpen
  \bibfield  {author} {\bibinfo {author} {\bibfnamefont {M.}~\bibnamefont
  {Smith}}, \bibinfo {author} {\bibfnamefont {A.~V.}\ \bibnamefont {Andreev}},
  \bibinfo {author} {\bibfnamefont {M.~V.}\ \bibnamefont {Feigel'man}}, \ and\
  \bibinfo {author} {\bibfnamefont {B.~Z.}\ \bibnamefont {Spivak}},\ }
  {\bibfield  {journal} {\bibinfo  {journal} {Physical Review B}\ }\textbf
  {\bibinfo {volume} {102}},\ \bibinfo {pages} {180507} (\bibinfo {year}
  {2020})}  \BibitemShut {NoStop}%
\bibitem [{\citenamefont {Kogan}\ and\ \citenamefont
  {Nakagawa}(2021)}]{Kogan2021a}%
  \BibitemOpen
  \bibfield  {author} {\bibinfo {author} {\bibfnamefont {V.~G.}\ \bibnamefont
  {Kogan}}\ and\ \bibinfo {author} {\bibfnamefont {N.}~\bibnamefont
  {Nakagawa}},\ } {\bibfield  {journal} {\bibinfo  {journal}
  {Physical Review B}\ }\textbf {\bibinfo {volume} {103}},\ \bibinfo {pages}
  {134511} (\bibinfo {year} {2021})}\BibitemShut {NoStop}%
\bibitem [{\citenamefont {Kasahara}\ \emph {et~al.}(2014)\citenamefont
  {Kasahara}, \citenamefont {Watashige}, \citenamefont {Hanaguri},
  \citenamefont {Kohsaka}, \citenamefont {Yamashita}, \citenamefont
  {Shimoyama}, \citenamefont {Mizukami}, \citenamefont {Endo}, \citenamefont
  {Ikeda}, \citenamefont {Aoyama}, \citenamefont {Terashima}, \citenamefont
  {Uji}, \citenamefont {Wolf}, \citenamefont {{Von L{\"{o}}hneysen}},
  \citenamefont {Shibauchi},\ and\ \citenamefont {Matsuda}}]{Kasahara2014}%
  \BibitemOpen
  \bibfield  {author} {\bibinfo {author} {\bibfnamefont {S.}~\bibnamefont
  {Kasahara}}, \bibinfo {author} {\bibfnamefont {T.}~\bibnamefont {Watashige}},
  \bibinfo {author} {\bibfnamefont {T.}~\bibnamefont {Hanaguri}}, \bibinfo
  {author} {\bibfnamefont {Y.}~\bibnamefont {Kohsaka}}, \bibinfo {author}
  {\bibfnamefont {T.}~\bibnamefont {Yamashita}}, \bibinfo {author}
  {\bibfnamefont {Y.}~\bibnamefont {Shimoyama}}, \bibinfo {author}
  {\bibfnamefont {Y.}~\bibnamefont {Mizukami}}, \bibinfo {author}
  {\bibfnamefont {R.}~\bibnamefont {Endo}}, \bibinfo {author} {\bibfnamefont
  {H.}~\bibnamefont {Ikeda}}, \bibinfo {author} {\bibfnamefont
  {K.}~\bibnamefont {Aoyama}}, \bibinfo {author} {\bibfnamefont
  {T.}~\bibnamefont {Terashima}}, \bibinfo {author} {\bibfnamefont
  {S.}~\bibnamefont {Uji}}, \bibinfo {author} {\bibfnamefont {T.}~\bibnamefont
  {Wolf}}, \bibinfo {author} {\bibfnamefont {H.}~\bibnamefont {{Von
  L{\"{o}}hneysen}}}, \bibinfo {author} {\bibfnamefont {T.}~\bibnamefont
  {Shibauchi}}, \ and\ \bibinfo {author} {\bibfnamefont {Y.}~\bibnamefont
  {Matsuda}},\ } {\bibfield  {journal} {\bibinfo  {journal}
  {Proceedings of the National Academy of Sciences of the United States of
  America}\ }\textbf {\bibinfo {volume} {111}},\ \bibinfo {pages} {16309}
  (\bibinfo {year} {2014})} \BibitemShut {NoStop}%
%\bibitem [{\citenamefont {Takahashi}\ \emph {et~al.}(2011)\citenamefont {Takahashi},
%  \citenamefont {Imai}, \citenamefont {Komiya}, \citenamefont {Tsukada},
%  \ and\ \citenamefont {Maeda}}]{Takahashi2011}%
%  \BibitemOpen
%  \bibfield  {author} {\bibinfo {author} {\bibfnamefont {H.}~\bibnamefont
%  {Takahashi}}, \bibinfo {author} {\bibfnamefont {Y.}~\bibnamefont {Imai}},
%  \bibinfo {author} {\bibfnamefont {S.}~\bibnamefont {Komiya}}, \bibinfo
%  {author} {\bibfnamefont {I.}~\bibnamefont {Tsukada}}, \ and\ \bibinfo {author}
%  {\bibfnamefont {A.}~\bibnamefont {Maeda}},\ }{\bibfield  {journal}
%  {\bibinfo  {journal} {Physical Review B}\
%  }\textbf {\bibinfo {volume} {84}},\ \bibinfo {pages} {132503} (\bibinfo
%  {year} {2011})}\BibitemShut {NoStop}%
\bibitem [{\citenamefont {Okada}\ \emph {et~al.}(2021)\citenamefont {Okada},
  \citenamefont {Imai}, \citenamefont {Urata}, \citenamefont {Tanabe},
  \citenamefont {Tanigaki},\ and\ \citenamefont {Maeda}}]{Okada2021}%
  \BibitemOpen
  \bibfield  {author} {\bibinfo {author} {\bibfnamefont {T.}~\bibnamefont
  {Okada}}, \bibinfo {author} {\bibfnamefont {Y.}~\bibnamefont {Imai}},
  \bibinfo {author} {\bibfnamefont {T.}~\bibnamefont {Urata}}, \bibinfo
  {author} {\bibfnamefont {Y.}~\bibnamefont {Tanabe}}, \bibinfo {author}
  {\bibfnamefont {K.}~\bibnamefont {Tanigaki}}, \ and\ \bibinfo {author}
  {\bibfnamefont {A.}~\bibnamefont {Maeda}},\ } {\bibfield  {journal}{\bibinfo  {journal} {Journal of the Physical Society of Japan}\
  }\textbf {\bibinfo {volume} {90}},\ \bibinfo {pages} {094704} (\bibinfo
  {year} {2021})}\BibitemShut {NoStop}%
% \bibitem [{\citenamefont {Kurokawa}\ \emph {et~al.}(2021)\citenamefont {Kurokawa},
%  \citenamefont {Nakamura}, \citenamefont {Zhao}, \citenamefont {Shikama},
%  \citenamefont {Sakishita},\citenamefont {Nabeshima},\citenamefont {Imai},\citenamefont {Kitano},\ and\ \citenamefont {Maeda}}]{Kurokawa2021}%
%  \BibitemOpen
%  \bibfield  {author} {\bibinfo {author} {\bibfnamefont {H.}~\bibnamefont
%  {Kurokawa}}, \bibinfo {author} {\bibfnamefont {S.}~\bibnamefont {Nakamura}},
%  \bibinfo {author} {\bibfnamefont {J.}~\bibnamefont {Zhao}}, \bibinfo
%  {author} {\bibfnamefont {N.}~\bibnamefont {Shikama}}, \bibinfo {author}
%  {\bibfnamefont {Y.}~\bibnamefont {Sakishita}}, \bibinfo {author} {\bibfnamefont
%  {Y.}~\bibnamefont {Sun}}, \bibinfo {author} {\bibfnamefont {F.}~\bibnamefont
%  {Nabeshima}}, \bibinfo {author} {\bibfnamefont {Y.}~\bibnamefont {Imai}} ,\bibinfo
%  {author} {\bibfnamefont {H.}~\bibnamefont {Kitano}},\ and\ \bibinfo {author}
%  {\bibfnamefont {A.}~\bibnamefont {Maeda}},\ } {\bibfield
%  {journal} {\bibinfo  {journal} {Physical Review B}\
%  }\textbf {\bibinfo {volume} {104}},\ \bibinfo {pages} {014505} (\bibinfo
%  {year} {2021})}\BibitemShut {NoStop}%
\bibitem [{\citenamefont {Hanaguri}\ \emph {et~al.}(2019)\citenamefont
  {Hanaguri}, \citenamefont {Kasahara}, \citenamefont {B{\"{o}}ker},
  \citenamefont {Eremin}, \citenamefont {Shibauchi},\ and\ \citenamefont
  {Matsuda}}]{Hanaguri2019}%
  \BibitemOpen
  \bibfield  {author} {\bibinfo {author} {\bibfnamefont {T.}~\bibnamefont
  {Hanaguri}}, \bibinfo {author} {\bibfnamefont {S.}~\bibnamefont {Kasahara}},
  \bibinfo {author} {\bibfnamefont {J.}~\bibnamefont {B{\"{o}}ker}}, \bibinfo
  {author} {\bibfnamefont {I.}~\bibnamefont {Eremin}}, \bibinfo {author}
  {\bibfnamefont {T.}~\bibnamefont {Shibauchi}}, \ and\ \bibinfo {author}
  {\bibfnamefont {Y.}~\bibnamefont {Matsuda}},\ }{\bibfield  {journal} {\bibinfo  {journal}
  {Physical Review Letters}\ }\textbf {\bibinfo {volume} {122}},\ \bibinfo
  {pages} {077001} (\bibinfo {year} {2019})} \BibitemShut {NoStop}%
\bibitem [{\citenamefont {Lin}\ and\ \citenamefont
  {Bulaevskii}(2013)}]{Lin2013}%
  \BibitemOpen
  \bibfield  {author} {\bibinfo {author} {\bibfnamefont {S.-Z.}\ \bibnamefont
  {Lin}}\ and\ \bibinfo {author} {\bibfnamefont {L.~N.}\ \bibnamefont
  {Bulaevskii}},\ }{\bibfield  {journal} {\bibinfo  {journal}
  {Physical Review Letters}\ }\textbf {\bibinfo {volume} {110}},\ \bibinfo
  {pages} {087003} (\bibinfo {year} {2013})}\BibitemShut {NoStop}%
\bibitem [{\citenamefont {Shibata}\ \emph {et~al.}(2003)\citenamefont
  {Shibata}, \citenamefont {Matsumoto}, \citenamefont {Izawa}, \citenamefont
  {Matsuda}, \citenamefont {Lee},\ and\ \citenamefont {Tajima}}]{Shibata2003}%
  \BibitemOpen
  \bibfield  {author} {\bibinfo {author} {\bibfnamefont {A.}~\bibnamefont
  {Shibata}}, \bibinfo {author} {\bibfnamefont {M.}~\bibnamefont {Matsumoto}},
  \bibinfo {author} {\bibfnamefont {K.}~\bibnamefont {Izawa}}, \bibinfo
  {author} {\bibfnamefont {Y.}~\bibnamefont {Matsuda}}, \bibinfo {author}
  {\bibfnamefont {S.}~\bibnamefont {Lee}}, \ and\ \bibinfo {author}
  {\bibfnamefont {S.}~\bibnamefont {Tajima}},\ }
  {\bibfield  {journal} {\bibinfo  {journal} {Physical Review B}\ }\textbf
  {\bibinfo {volume} {68}},\ \bibinfo {pages} {060501} (\bibinfo {year}
  {2003})}\BibitemShut {NoStop}%
\bibitem [{\citenamefont {Akutagawa}\ \emph {et~al.}(2008)\citenamefont
  {Akutagawa}, \citenamefont {Ohashi}, \citenamefont {Kitano}, \citenamefont
  {Maeda}, \citenamefont {Goryo}, \citenamefont {Matsukawa},\ and\
  \citenamefont {Akimitsu}}]{Akutagawa2008}%
  \BibitemOpen
  \bibfield  {author} {\bibinfo {author} {\bibfnamefont {S.}~\bibnamefont
  {Akutagawa}}, \bibinfo {author} {\bibfnamefont {T.}~\bibnamefont {Ohashi}},
  \bibinfo {author} {\bibfnamefont {H.}~\bibnamefont {Kitano}}, \bibinfo
  {author} {\bibfnamefont {A.}~\bibnamefont {Maeda}}, \bibinfo {author}
  {\bibfnamefont {J.}~\bibnamefont {Goryo}}, \bibinfo {author} {\bibfnamefont
  {H.}~\bibnamefont {Matsukawa}}, \ and\ \bibinfo {author} {\bibfnamefont
  {J.}~\bibnamefont {Akimitsu}},\ } {\bibfield  {journal} {\bibinfo
  {journal} {Journal of the Physical Society of Japan}\ }\textbf {\bibinfo
  {volume} {77}},\ \bibinfo {pages} {064701} (\bibinfo {year}
  {2008})}\BibitemShut {NoStop}%
\bibitem [{\citenamefont {Okada}\ \emph {et~al.}(2012)\citenamefont {Okada},
  \citenamefont {Takahashi}, \citenamefont {Imai}, \citenamefont {Kitagawa},
  \citenamefont {Matsubayashi}, \citenamefont {Uwatoko},\ and\ \citenamefont
  {Maeda}}]{Okada2012}%
  \BibitemOpen
  \bibfield  {author} {\bibinfo {author} {\bibfnamefont {T.}~\bibnamefont
  {Okada}}, \bibinfo {author} {\bibfnamefont {H.}~\bibnamefont {Takahashi}},
  \bibinfo {author} {\bibfnamefont {Y.}~\bibnamefont {Imai}}, \bibinfo {author}
  {\bibfnamefont {K.}~\bibnamefont {Kitagawa}}, \bibinfo {author}
  {\bibfnamefont {K.}~\bibnamefont {Matsubayashi}}, \bibinfo {author}
  {\bibfnamefont {Y.}~\bibnamefont {Uwatoko}}, \ and\ \bibinfo {author}
  {\bibfnamefont {A.}~\bibnamefont {Maeda}},\ }
  {\bibfield  {journal} {\bibinfo  {journal} {Physical Review B}\ }\textbf
  {\bibinfo {volume} {86}},\ \bibinfo {pages} {064516} (\bibinfo {year}
  {2012})} \BibitemShut {NoStop}%
\bibitem [{\citenamefont {Takahashi}\ \emph {et~al.}(2012)\citenamefont
  {Takahashi}, \citenamefont {Okada}, \citenamefont {Imai}, \citenamefont
  {Kitagawa}, \citenamefont {Matsubayashi}, \citenamefont {Uwatoko},\ and\
  \citenamefont {Maeda}}]{Takahashi2012a}%
  \BibitemOpen
  \bibfield  {author} {\bibinfo {author} {\bibfnamefont {H.}~\bibnamefont
  {Takahashi}}, \bibinfo {author} {\bibfnamefont {T.}~\bibnamefont {Okada}},
  \bibinfo {author} {\bibfnamefont {Y.}~\bibnamefont {Imai}}, \bibinfo {author}
  {\bibfnamefont {K.}~\bibnamefont {Kitagawa}}, \bibinfo {author}
  {\bibfnamefont {K.}~\bibnamefont {Matsubayashi}}, \bibinfo {author}
  {\bibfnamefont {Y.}~\bibnamefont {Uwatoko}}, \ and\ \bibinfo {author}
  {\bibfnamefont {A.}~\bibnamefont {Maeda}},\ }
  {\bibfield  {journal} {\bibinfo  {journal} {Physical Review B}\ }\textbf
  {\bibinfo {volume} {86}},\ \bibinfo {pages} {144525} (\bibinfo {year}
  {2012})} \BibitemShut {NoStop}%
\bibitem [{\citenamefont {Okada}\ \emph
  {et~al.}(2013{\natexlab{a}})\citenamefont {Okada}, \citenamefont {Takahashi},
  \citenamefont {Imai}, \citenamefont {Kitagawa}, \citenamefont {Matsubayashi},
  \citenamefont {Uwatoko},\ and\ \citenamefont {Maeda}}]{Okada2013a}%
  \BibitemOpen
  \bibfield  {author} {\bibinfo {author} {\bibfnamefont {T.}~\bibnamefont
  {Okada}}, \bibinfo {author} {\bibfnamefont {H.}~\bibnamefont {Takahashi}},
  \bibinfo {author} {\bibfnamefont {Y.}~\bibnamefont {Imai}}, \bibinfo {author}
  {\bibfnamefont {K.}~\bibnamefont {Kitagawa}}, \bibinfo {author}
  {\bibfnamefont {K.}~\bibnamefont {Matsubayashi}}, \bibinfo {author}
  {\bibfnamefont {Y.}~\bibnamefont {Uwatoko}}, \ and\ \bibinfo {author}
  {\bibfnamefont {A.}~\bibnamefont {Maeda}},\ }{\bibfield  {journal} {\bibinfo
  {journal} {Physica C: Superconductivity}\ }\textbf {\bibinfo {volume}
  {484}},\ \bibinfo {pages} {27} (\bibinfo {year}
  {2013}{\natexlab{a}})}\BibitemShut {NoStop}%
\bibitem [{\citenamefont {Okada}\ \emph
  {et~al.}(2013{\natexlab{b}})\citenamefont {Okada}, \citenamefont {Takahashi},
  \citenamefont {Imai}, \citenamefont {Kitagawa}, \citenamefont {Matsubayashi},
  \citenamefont {Uwatoko},\ and\ \citenamefont {Maeda}}]{Okada2013b}%
  \BibitemOpen
  \bibfield  {author} {\bibinfo {author} {\bibfnamefont {T.}~\bibnamefont
  {Okada}}, \bibinfo {author} {\bibfnamefont {H.}~\bibnamefont {Takahashi}},
  \bibinfo {author} {\bibfnamefont {Y.}~\bibnamefont {Imai}}, \bibinfo {author}
  {\bibfnamefont {K.}~\bibnamefont {Kitagawa}}, \bibinfo {author}
  {\bibfnamefont {K.}~\bibnamefont {Matsubayashi}}, \bibinfo {author}
  {\bibfnamefont {Y.}~\bibnamefont {Uwatoko}}, \ and\ \bibinfo {author}
  {\bibfnamefont {A.}~\bibnamefont {Maeda}},\ }{\bibfield  {journal} {\bibinfo
  {journal} {Physica C: Superconductivity}\ }\textbf {\bibinfo {volume}
  {494}},\ \bibinfo {pages} {109} (\bibinfo {year}
  {2013}{\natexlab{b}})}\BibitemShut {NoStop}%
\bibitem [{\citenamefont {Okada}\ \emph {et~al.}(2014)\citenamefont {Okada},
  \citenamefont {Imai}, \citenamefont {Takahashi}, \citenamefont {Nakajima},
  \citenamefont {Iyo}, \citenamefont {Eisaki},\ and\ \citenamefont
  {Maeda}}]{Okada2014}%
  \BibitemOpen
  \bibfield  {author} {\bibinfo {author} {\bibfnamefont {T.}~\bibnamefont
  {Okada}}, \bibinfo {author} {\bibfnamefont {Y.}~\bibnamefont {Imai}},
  \bibinfo {author} {\bibfnamefont {H.}~\bibnamefont {Takahashi}}, \bibinfo
  {author} {\bibfnamefont {M.}~\bibnamefont {Nakajima}}, \bibinfo {author}
  {\bibfnamefont {A.}~\bibnamefont {Iyo}}, \bibinfo {author} {\bibfnamefont
  {H.}~\bibnamefont {Eisaki}}, \ and\ \bibinfo {author} {\bibfnamefont
  {A.}~\bibnamefont {Maeda}},\ } {\bibfield  {journal} {\bibinfo  {journal}
  {Physica C: Superconductivity and its Applications}\ }\textbf {\bibinfo
  {volume} {504}},\ \bibinfo {pages} {24} (\bibinfo {year} {2014})}\BibitemShut
  {NoStop}%
\bibitem [{\citenamefont {Okada}\ \emph {et~al.}(2015)\citenamefont {Okada},
  \citenamefont {Nabeshima}, \citenamefont {Takahashi}, \citenamefont {Imai},\
  and\ \citenamefont {Maeda}}]{Okada2015}%
  \BibitemOpen
  \bibfield  {author} {\bibinfo {author} {\bibfnamefont {T.}~\bibnamefont
  {Okada}}, \bibinfo {author} {\bibfnamefont {F.}~\bibnamefont {Nabeshima}},
  \bibinfo {author} {\bibfnamefont {H.}~\bibnamefont {Takahashi}}, \bibinfo
  {author} {\bibfnamefont {Y.}~\bibnamefont {Imai}}, \ and\ \bibinfo {author}
  {\bibfnamefont {A.}~\bibnamefont {Maeda}},\ }{\bibfield  {journal} {\bibinfo
   {journal} {Physical Review B}\ }\textbf {\bibinfo {volume} {91}},\ \bibinfo
  {pages} {054510} (\bibinfo {year} {2015})}\BibitemShut {NoStop}%
\bibitem [{\citenamefont {B{\"{o}}hmer}\ \emph {et~al.}(2013)\citenamefont
  {B{\"{o}}hmer}, \citenamefont {Hardy}, \citenamefont {Eilers}, \citenamefont
  {Ernst}, \citenamefont {Adelmann}, \citenamefont {Schweiss}, \citenamefont
  {Wolf},\ and\ \citenamefont {Meingast}}]{Bohmer2013}%
  \BibitemOpen
  \bibfield  {author} {\bibinfo {author} {\bibfnamefont {A.~E.}\ \bibnamefont
  {B{\"{o}}hmer}}, \bibinfo {author} {\bibfnamefont {F.}~\bibnamefont {Hardy}},
  \bibinfo {author} {\bibfnamefont {F.}~\bibnamefont {Eilers}}, \bibinfo
  {author} {\bibfnamefont {D.}~\bibnamefont {Ernst}}, \bibinfo {author}
  {\bibfnamefont {P.}~\bibnamefont {Adelmann}}, \bibinfo {author}
  {\bibfnamefont {P.}~\bibnamefont {Schweiss}}, \bibinfo {author}
  {\bibfnamefont {T.}~\bibnamefont {Wolf}}, \ and\ \bibinfo {author}
  {\bibfnamefont {C.}~\bibnamefont {Meingast}},\ }{\bibfield  {journal} {\bibinfo
  {journal} {Physical Review B}\ }\textbf {\bibinfo {volume} {87}},\ \bibinfo
  {pages} {180505} (\bibinfo {year} {2013})}\BibitemShut {NoStop}%
\bibitem [{\citenamefont {Lei}\ \emph {et~al.}(2012)\citenamefont {Lei},
  \citenamefont {Graf}, \citenamefont {Hu}, \citenamefont {Ryu}, \citenamefont
  {Choi}, \citenamefont {Tozer},\ and\ \citenamefont {Petrovic}}]{Lei2012}%
  \BibitemOpen
  \bibfield  {author} {\bibinfo {author} {\bibfnamefont {H.}~\bibnamefont
  {Lei}}, \bibinfo {author} {\bibfnamefont {D.}~\bibnamefont {Graf}}, \bibinfo
  {author} {\bibfnamefont {R.}~\bibnamefont {Hu}}, \bibinfo {author}
  {\bibfnamefont {H.}~\bibnamefont {Ryu}}, \bibinfo {author} {\bibfnamefont
  {E.~S.}\ \bibnamefont {Choi}}, \bibinfo {author} {\bibfnamefont {S.~W.}\
  \bibnamefont {Tozer}}, \ and\ \bibinfo {author} {\bibfnamefont
  {C.}~\bibnamefont {Petrovic}},\ } {\bibfield  {journal} {\bibinfo  {journal}
  {Physical Review B}\ }\textbf {\bibinfo {volume} {85}},\ \bibinfo {pages}
  {094515} (\bibinfo {year} {2012})}\BibitemShut {NoStop}%
\bibitem [{\citenamefont {Subedi}\ \emph {et~al.}(2008)\citenamefont {Subedi},
  \citenamefont {Zhang}, \citenamefont {Singh},\ and\ \citenamefont
  {Du}}]{Subedi2008}%
  \BibitemOpen
  \bibfield  {author} {\bibinfo {author} {\bibfnamefont {A.}~\bibnamefont
  {Subedi}}, \bibinfo {author} {\bibfnamefont {L.}~\bibnamefont {Zhang}},
  \bibinfo {author} {\bibfnamefont {D.~J.}\ \bibnamefont {Singh}}, \ and\
  \bibinfo {author} {\bibfnamefont {M.~H.}\ \bibnamefont {Du}},\ }{\bibfield  {journal} {\bibinfo
   {journal} {Physical Review B}\ }\textbf {\bibinfo {volume} {78}},\ \bibinfo
  {pages} {134514} (\bibinfo {year} {2008})}\BibitemShut {NoStop}%
\bibitem [{\citenamefont {Jin}\ \emph {et~al.}(2001)\citenamefont {Jin},
  \citenamefont {Paranthaman}, \citenamefont {Zhai}, \citenamefont {Christen},
  \citenamefont {Christen},\ and\ \citenamefont {Mandrus}}]{Jin2001}%
  \BibitemOpen
  \bibfield  {author} {\bibinfo {author} {\bibfnamefont {R.}~\bibnamefont
  {Jin}}, \bibinfo {author} {\bibfnamefont {M.}~\bibnamefont {Paranthaman}},
  \bibinfo {author} {\bibfnamefont {H.~Y.}\ \bibnamefont {Zhai}}, \bibinfo
  {author} {\bibfnamefont {H.~M.}\ \bibnamefont {Christen}}, \bibinfo {author}
  {\bibfnamefont {D.~K.}\ \bibnamefont {Christen}}, \ and\ \bibinfo {author}
  {\bibfnamefont {D.}~\bibnamefont {Mandrus}},\ }{\bibfield  {journal} {\bibinfo
   {journal} {Physical Review B}\ }\textbf {\bibinfo {volume} {64}},\ \bibinfo
  {pages} {220506} (\bibinfo {year} {2001})}\BibitemShut {NoStop}%
\bibitem [{\citenamefont {Lei}\ \emph {et~al.}(2010)\citenamefont {Lei},
  \citenamefont {Hu}, \citenamefont {Choi},\ and\ \citenamefont
  {Petrovic}}]{Lei2010}%
  \BibitemOpen
  \bibfield  {author} {\bibinfo {author} {\bibfnamefont {H.}~\bibnamefont
  {Lei}}, \bibinfo {author} {\bibfnamefont {R.}~\bibnamefont {Hu}}, \bibinfo
  {author} {\bibfnamefont {E.~S.}\ \bibnamefont {Choi}}, \ and\ \bibinfo
  {author} {\bibfnamefont {C.}~\bibnamefont {Petrovic}},\ }{\bibfield  {journal} {\bibinfo  {journal}
  {Physical Review B}\ }\textbf {\bibinfo {volume} {82}},\ \bibinfo {pages}
  {134525} (\bibinfo {year} {2010})}\BibitemShut {NoStop}%
\bibitem [{\citenamefont {Wang}\ \emph {et~al.}(2011)\citenamefont {Wang},
  \citenamefont {Sou}, \citenamefont {Yang}, \citenamefont {Chang},
  \citenamefont {Cheng}, \citenamefont {Tsuei}, \citenamefont {Su},
  \citenamefont {Wolf},\ and\ \citenamefont {Adelmann}}]{Wang2011}%
  \BibitemOpen
  \bibfield  {author} {\bibinfo {author} {\bibfnamefont {L.~M.}\ \bibnamefont
  {Wang}}, \bibinfo {author} {\bibfnamefont {U.-C.}\ \bibnamefont {Sou}},
  \bibinfo {author} {\bibfnamefont {H.~C.}\ \bibnamefont {Yang}}, \bibinfo
  {author} {\bibfnamefont {L.~J.}\ \bibnamefont {Chang}}, \bibinfo {author}
  {\bibfnamefont {C.-M.}\ \bibnamefont {Cheng}}, \bibinfo {author}
  {\bibfnamefont {K.-D.}\ \bibnamefont {Tsuei}}, \bibinfo {author}
  {\bibfnamefont {Y.}~\bibnamefont {Su}}, \bibinfo {author} {\bibfnamefont
  {T.}~\bibnamefont {Wolf}}, \ and\ \bibinfo {author} {\bibfnamefont
  {P.}~\bibnamefont {Adelmann}},\ }{\bibfield  {journal} {\bibinfo  {journal}
  {Physical Review B}\ }\textbf {\bibinfo {volume} {83}},\ \bibinfo {pages}
  {134506} (\bibinfo {year} {2011})}\BibitemShut {NoStop}%
\bibitem [{\citenamefont {Babaev}(2002)}]{Babaev2002}%
  \BibitemOpen
  \bibfield  {author} {\bibinfo {author} {\bibfnamefont {E.}~\bibnamefont
  {Babaev}},\ } {\bibfield  {journal}
  {\bibinfo  {journal} {Physical Review Letters}\ }\textbf {\bibinfo {volume}
  {89}},\ \bibinfo {pages} {067001} (\bibinfo {year} {2002})}\BibitemShut
  {NoStop}%
\bibitem [{\citenamefont {Silaev}\ and\ \citenamefont
  {Vargunin}(2016)}]{Silaev2016}%
  \BibitemOpen
  \bibfield  {author} {\bibinfo {author} {\bibfnamefont {M.}~\bibnamefont
  {Silaev}}\ and\ \bibinfo {author} {\bibfnamefont {A.}~\bibnamefont
  {Vargunin}},\ } {\bibfield  {journal} {\bibinfo
  {journal} {Physical Review B}\ }\textbf {\bibinfo {volume} {94}},\ \bibinfo
  {pages} {1} (\bibinfo {year} {2016})}\BibitemShut {NoStop}%
 \bibitem [{\citenamefont {Kopnin}(2001)}]{Kopnin2001}%
  \BibitemOpen
  \bibfield  {author} {\bibinfo {author} {\bibfnamefont {N.~B.}\ \bibnamefont
  {Kopnin}},\ }\href@noop {} {\emph {\bibinfo {title} {{Theory of
  nonequilibrium superconductivity}}}},\ Vol.\ \bibinfo {volume} {110}\
  (\bibinfo  {publisher} {Oxford University Press.},\ \bibinfo {year}
  {2001})\BibitemShut {NoStop}%
 \bibitem [{\citenamefont {Liu}\ and\ \citenamefont
  {Liu}(2018)}]{Liu2018}%
  \BibitemOpen
  \bibfield  {author} {\bibinfo {author} {\bibfnamefont {D.}~\bibnamefont
  {Liu}}, \bibinfo {author} {\bibfnamefont {C.}~\bibnamefont {Li}}, \bibinfo
  {author} {\bibfnamefont {J.}~\bibnamefont {Huang}}, \bibinfo
  {author} {\bibfnamefont {B.}~\bibnamefont {Lei}}, \bibinfo {author} {\bibfnamefont {L.}~\bibnamefont {Wang}}, \bibinfo {author} {\bibfnamefont {X.}~\bibnamefont {Wu}}, \bibinfo {author} {\bibfnamefont {B.}~\bibnamefont {Shen}}, \bibinfo {author} {\bibfnamefont {Q.}~\bibnamefont {Gao}},\bibinfo {author} {\bibfnamefont {Y.}~\bibnamefont {Zhang}},  \bibinfo {author} {\bibfnamefont {X.}~\bibnamefont {Liu}}, \bibinfo {author} {\bibfnamefont {Y.}~\bibnamefont {Hu}}, \bibinfo {author} {\bibfnamefont {A.}~\bibnamefont {Liang}}, \bibinfo {author} {\bibfnamefont {J.}~\bibnamefont {Liu}}, \bibinfo {author} {\bibfnamefont {P.}~\bibnamefont {Ai}},\bibinfo {author} {\bibfnamefont {L.}~\bibnamefont {Zhao}},  \bibinfo {author} {\bibfnamefont {S.}~\bibnamefont {He}}, \bibinfo {author} {\bibfnamefont {L.}~\bibnamefont {Yu}},\bibinfo {author} {\bibfnamefont {G.}~\bibnamefont {Liu}},  \bibinfo {author} {\bibfnamefont {Y.}~\bibnamefont {Mao}}, \bibinfo {author} {\bibfnamefont {X.}~\bibnamefont {Dong}}, \bibinfo {author} {\bibfnamefont {X.}~\bibnamefont {Jia}}, \bibinfo {author} {\bibfnamefont {F.}~\bibnamefont {Zhang}}, \bibinfo {author} {\bibfnamefont {F.}~\bibnamefont {Zhang}}, \bibinfo {author} {\bibfnamefont {F.}~\bibnamefont {Yang}},  \bibinfo {author} {\bibfnamefont {Z.}~\bibnamefont {Wang}}, \bibinfo {author} {\bibfnamefont {Q.}~\bibnamefont {Peng}}, \bibinfo {author} {\bibfnamefont {Y.}~\bibnamefont {Shi}}, \bibinfo {author} {\bibfnamefont {J.}~\bibnamefont {Hu}}, \bibinfo {author} {\bibfnamefont {T.}~\bibnamefont {Xiang}}, \bibinfo {author} {\bibfnamefont {X.}~\bibnamefont {Chen}}, \bibinfo {author} {\bibfnamefont {Z.}~\bibnamefont {Wu}}, \bibinfo {author} {\bibfnamefont {C.}~\bibnamefont {Chen}},  and\ \bibinfo {author} {\bibfnamefont {X. J.}~\bibnamefont {Zhou}},\ } {\bibfield  {journal} {\bibinfo  {journal}
  {Physical Review X}\ }\textbf {\bibinfo {volume} {8}},\ \bibinfo {pages}
  {031033} (\bibinfo {year} {2018})}\BibitemShut {NoStop}%
  \bibitem [{\citenamefont {Okada}\ and\ \citenamefont
  {Okada}(2020)}]{Okada2020}%
  \BibitemOpen
  \bibfield  {author} {\bibinfo {author} {\bibfnamefont {T.}~\bibnamefont
  {Okada}}, \bibinfo {author} {\bibfnamefont {Y.}~\bibnamefont {Imai}}, \bibinfo
  {author} {\bibfnamefont {K.}~\bibnamefont {Kitagawa}}, \bibinfo
  {author} {\bibfnamefont {K.}~\bibnamefont {Matsubayashi}}, \bibinfo {author} {\bibfnamefont {M.}~\bibnamefont {Nakajima}}, \bibinfo {author} {\bibfnamefont {A.}~\bibnamefont {Iyo}}, \bibinfo {author} {\bibfnamefont {Y.}~\bibnamefont {Uwatoko}}, \bibinfo {author} {\bibfnamefont {H.}~\bibnamefont {Eisaki}}, and\ \bibinfo {author} {\bibfnamefont {A.}~\bibnamefont {Maeda}},\ } {\bibfield  {journal} {\bibinfo  {journal}
  {Sci Rep}\ }\textbf {\bibinfo {volume} {10}},\ \bibinfo {pages}
  {7064} (\bibinfo {year} {2020})}\BibitemShut {NoStop}%
\end{thebibliography}
\end{document}